\documentclass{llncs}

\usepackage{balance} 
\usepackage{paper-2024-emas-concurrency}
\usepackage[inline]{enumitem}

\acrodef{1A1P}{One-Agent-One-Process}
\acrodef{1A1T}{One-Agent-One-Thread}
\acrodef{AA1E}{All-Agents-One-Executor}
\acrodef{AA1EL}{All-Agents-One-Event-Loop}
\acrodef{AA1T}{All-Agents-One-Thread}
\acrodef{AOP}{Agent-Oriented Programming}
\acrodef{API}{Application Programming Interface}
\acrodef{BDI}{Belief-Desire-Intention}
\acrodef{CCS}{Calculus of Communicating Systems}
\acrodef{DSL}{Domain-Specific Language}
\acrodef{FIFO}{First-In-First-Out}
\acrodef{GIL}{Global Interpreter Lock}
\acrodef{JNI}{Java Native Interface}
\acrodef{JVM}{Java Virtual Machine}
\acrodef{MAS}{Multi-Agent System}
\acrodef{OS}{Operating System}

\begin{document}

\title{On the External Concurrency\\of Current BDI Frameworks for MAS}

\titlerunning{External Concurrency of Current BDI Frameworks}

\authorrunning{Baiardi et al.}

\author{Martina Baiardi\inst{1}\orcidID{0009-0001-0799-9166}\and
Samuele Burattini\inst{1}\orcidID{0009-0009-4853-7783}\and
Giovanni Ciatto\inst{1}\orcidID{0000-0002-1841-8996} \and
Danilo Pianini\inst{1}\orcidID{0000-0002-8392-5409} \and
Alessandro Ricci\inst{1}\orcidID{0000-0002-9222-5092} \and
Andrea Omicini\inst{1}\orcidID{0000-0002-6655-3869}
}

\institute{
    \disi{} (DISI)
    \\
    \unibo
    \\
    Via dell'Università 50, 47522 Cesena (FC), Italy
    \\\smallskip
    \email{ $\{ m.baiardi, samuele.burattini, giovanni.ciatto,\allowbreak danilo.pianini, a.ricci, andrea.omicini\}$@unibo.it},
    \texttt{https://www.unibo.it/sitoweb/$\{ m.baiardi, samuele.burattini,\allowbreak giovanni.ciatto, danilo.pianini, a.ricci, andrea.omicini\}$/en}
}

\maketitle


\begin{abstract}
The execution of \ac{BDI} agents in a \ac{MAS}
can be practically implemented on top of low-level concurrency mechanisms
that impact on efficiency, determinism, and reproducibility.
We argue that developers should specify the \ac{MAS} behaviour independently of the execution model,
and choose or configure
the concurrency model later on,
according to the specific needs of their target domain,
leaving the  \ac{MAS} specification unaffected.
We identify patterns for mapping the agent execution over the underlying concurrency abstractions,
and investigate which concurrency models are supported by some of the most commonly used \ac{BDI} platforms.
Although most frameworks support multiple concurrency models,
we find that they mostly hide them under the hood,
making them opaque to the developer,
and actually limiting the possibility of fine-tuning the \ac{MAS}\@.


\end{abstract}


\section{Introduction}

The \ac{AOP} paradigm was introduced almost thirty years ago~\cite{shoham1993agent}
as a way to model software in terms of autonomous computational entities
capable of carrying on several courses of action \emph{simultaneously}---there including,
interacting with their environment and among each other.
Since its conception,
\ac{AOP} has been strictly linked with the \emph{strong} notion of agency~\cite{wooldridge1995intelligent-agents},
where agents are assumed to be aware of their own goals and able to reason about if,
when, and how to pursue them---not necessarily in a predefined order.
Along this line, \ac{AOP} frameworks evolved to embrace the \acf{BDI} model~\cite{bratman1988bdi},
where agents are modelled and implemented by means of abstractions mimicking typically human-level notions.
By construction,
\ac{BDI} agents are able to carry on multiple intentions at any given time~\cite{Rao1996},
and many research and software-development efforts have been devoted to the definition of \ac{BDI}
architectures and programming languages giving precise semantics
to the \emph{concurrent} execution of such intentions~\cite{BordiniHW2007}.

%
As computational entities, agents are autonomous as they encapsulate their own \emph{control flow}~\cite{objag-jot1}.
Control-flow encapsulation is commonly referred to as \emph{computational} autonomy~\cite{artifacts-jaamas17},
and it is considered a necessary -- yet not sufficient -- pre-requisite for autonomy in software agents.

On mainstream programming platforms
-- such as the \ac{JVM}~\cite{lptech4mas-jaamas35}, used for the implementation of many \ac{BDI} frameworks --,
computational autonomy can be achieved by mapping each agent onto a control-flow-related primitive:
a thread, a process, or an event loop.
This, in turn, enables and constraints the ways by which multiple agents may be \emph{concurrently} executed.
In this paper we refer to the mapping between \ac{BDI} abstractions and the underlying concurrency primitive
as the \emph{concurrency model} of the framework.

The selection of an appropriate concurrency model deeply impacts several aspects of the agent programming framework:
efficiency, determinism, and reproducibility.
In particular,
the concurrency model determines whether, and to what extent
\begin{inlinelist}
    \item multiple agents can run in parallel, and
    \item one agent can carry on parallel activities.
\end{inlinelist}
Parallelism, in turn, affects the efficiency of \ac{MAS} execution
(particularly on hardware supporting true parallel execution)
and the determinism of the overall \ac{MAS} dynamics.
In fact,
parallelism introduces non-deterministic interleaving of the agent's actions,
undermining predictability and reproducibility,
which may be a strict requirement in some applications,
such as multi-agent based simulation~\cite{BandiniMV09}.
Finely capturing and controlling concurrency is crucial in modern software engineering,
even beyond \acp{MAS}:
consider, for instance, trends such as event-driven~\cite{DabekZKMM2002} and reactive~\cite{bainomugisha2013survey} programming.

Unfortunately,
dealing with concurrency is commonly acknowledged as error-prone and challenging.
Thus,
mainstream programming languages and platforms are featuring more and more constructs
helping developers to leverage concurrency through better abstractions
(e.g., Javascript's async/await~\cite{LoringML17},
Akka's reactive streams~\cite{DBLP:journals/cai/DadelZ16},
and Kotlin's coroutines~\cite{DBLP:conf/oopsla/ElizarovBAU21}),
hiding part of the subtle intricacies under the hood.
\ac{AOP} tools and frameworks are no exception to this trend:
they come with one or more concurrency models,
often (in compliance with the information hiding principle)
hidden under the hood
to let programmers focus on the agents' behaviour.

In this work we argue that, although the separation of concurrency models and \ac{MAS} specifications is paramount,
removing control from developers' hand is not the best solution:
they should be aware of available possibilities and related trade-offs,
and select (and, possibly, swap) them depending on the
specific needs of their application and execution environment.
This is particularly true for \ac{BDI} agent technologies,
where the semantics of intention scheduling may be realised in many different ways.

\paragraph{Contribution.}
In this work, we introduce the notions of \emph{internal} and \emph{external} concurrency,
capturing, respectively, the concurrency among agents' activities and
the concurrency induced by the selection of the mapping of multiple agents onto the underlying concurrency abstractions.
These two abstractions influence each other:
enforcing one restricts the range of possibilities of the other,
impacting performance, determinism, and reproducibility.
Despite that, the previous literature focuses on internal concurrency,
leaving the external one as an implicit consequence of the choices
made to support internal behaviour.
Thus,
in this paper we provide a taxonomy of concurrency models that may be adopted by \ac{BDI} frameworks,
and we classify several notable \ac{BDI} agent technologies accordingly.
Finally, we draw practical engineering recommendations for the development of \ac{BDI} agent technologies,
suggesting to take into account the control of \emph{external} concurrency at design time.

\paragraph{Structure.}
The remainder of this paper is structured as follows.
In \Cref{sec:background}
we define internal and external concurrency in BDI agents,
and discuss how they have been considered in related works in the \ac{AOP} community.
We analyse concurrency models commonly adopted in modern software development,
then (\Cref{sec:taxonomy}) we discuss how agents (and their internal components) can be mapped onto them,
evaluating the pros and cons.
In \Cref{sec:comparison} we evaluate several \ac{BDI} technologies from the \ac{AOP} community from a concurrency-related perspective,
eliciting the available concurrency models and their degree of configurability.
Finally,
in \Cref{sec:discussion} we elaborate on the importance of configurable concurrency models
well-separated from the agent's behaviour specification.

\section{Background}\label{sec:background}

In this section
we first recall basic notions frame the concepts of \emph{internal} and \emph{external} concurrency,
then
look at the existing work specifically addressing concurrency in the context of \ac{BDI} \ac{AOP},
thus framing our contribution with respect to the state of the art.
Then
we discuss the lower-level concurrency abstractions required to understand the remainder of the paper.

\subsection{Internal vs.\ External Concurrency}

A multi-\ac{BDI}-agent system can be modelled in \ac{CCS}~\cite{Milner80} as a set of agents running in parallel.
Each agent is essentially an infinite loop where, at each iteration step, the three main stages of the agent's control loop are executed%
---sensing, deliberating, and acting.
More formally:
\begin{equation}\label{eq:external_concurrency}
\begin{array}{rcl}
  \V{mas} & \defeq & \V{agent}_1 \parallel \ldots \parallel \V{agent}_N \\
  \V{agent} & \defeq & \T{sense} \cdot \T{deliberate} \cdot \T{act} \cdot \V{agent} \\
\end{array}
\end{equation}
where
\begin{inlinelist}
  \item operation $\T{sense}$ handles new percepts and incoming messages, generating update events accordingly,
  \item operation $\T{deliberate}$ chooses how to handle those events and picks the next action to be executed
  and
  \item operation $\T{act}$ executes the selected action%
  ---e.g. sending a message, affecting the environment, or changing the agent's internal state.
\end{inlinelist}

This simple modelling focuses on the control loop of agents,
while hiding another key aspect of \ac{MAS}: interaction among agents%
---i.e. how each agent's actions may influence other agents.
Interaction may consist of either communication 
(e.g. direct message passing) 
or stigmergy 
(e.g. indirectly altering the environment to affect other agents).
In both cases, 
interaction implies one agent acting and another agent perceiving the effects of that action,
so, as far as concurrency and control-loops are concerned,
the modelling above is sufficient.

\subsubsection{Internal Concurrency} is the way in which those operations are modelled,
there including whether they are further decomposable or not, their degree of concurrency, and their interleaving.
For instance, in~\cite{ZatelliRH2015}, two major patterns are identified:
the \emph{synchronous} one,  where \emph{all} percepts and messages are \emph{sequentially} handled in the sensing stage,
and \emph{only one} action is selected by the deliberation stage and executed by the action stage:
\begin{equation}\label{eq:sequentialIC}
  \begin{array}{rcl}
    \V{agent} & \defeq & \V{sense} \cdot \V{deliberate} \cdot \V{Act} \cdot \V{agent} \\
    \V{sense} & \defeq & \T{sense}_1 \cdot \ldots \cdot \T{sense}_M \\
    \V{deliberate} & \defeq & \T{deliberate} \\
    \V{act} & \defeq & \T{act} \\
  \end{array}
\end{equation}
and the \emph{asynchronous} one, where \emph{multiple} percepts and messages are \emph{concurrently} handled in the sensing stage,
and deliberation and action stages are executed concurrently as well:
\begin{equation}\label{eq:parallelIC}
  \begin{array}{rcl}
    \V{agent} & \defeq & \V{sense} \parallel \V{deliberate} \parallel \V{act} \\
    \V{sense} & \defeq & (\T{sense}_1 \parallel \ldots \parallel \T{sense}_M) \cdot \V{sense} \\
    \V{deliberate} & \defeq & (\T{deliberate}_1 \parallel \ldots \parallel \T{deliberate}_L) \cdot \V{deliberate} \\
    \V{act} & \defeq & (\T{act}_1 \parallel \ldots \parallel \T{act}_K) \cdot \V{act} \\
  \end{array}
\end{equation}
Other patterns may be defined in this framework;
e.g., the single step of the control-loop can be modelled as a fork/join,
where all percepts are handled concurrently,
then all deliberations are handled concurrently, and then all actions are executed concurrently.
Yet, the key point is that all models focus on the execution of the agent's control loop,
and, by extension, on the interleaving of the agent's intentions.
For instance, 
a system modelled as in \Cref{eq:sequentialIC} would only support \emph{simulated} parallelism%
---e.g., a very common implementation is: 
each cycle of the control-loop executes a single action from a single intention.
Conversely, a system modelled as in \Cref{eq:parallelIC} would support \emph{true} parallelism%
---so, in principle, two or more action could be executed in the same moment.

\subsubsection{External Concurrency,} conversely, is what we focus on in this paper---i.e., the way the control loops of multiple agents are mapped onto the underlying concurrency abstractions
(\Cref{sec:concurrency_abstractions}).
In other words, we are interested in understanding how \Cref{eq:external_concurrency} can be -- and commonly is -- implemented in practice.
Arguably,
understanding and explicitly modelling external concurrency is crucial,
as the external concurrency model constrains and supports the admissible internal concurrency models:
the relationship between the two is bi-directional.
Also,
we argue that the external concurrency model has the most impact on the overall system properties.
For instance,
even a massively-parallel agent internal concurrency model would not lead to any speedup 
compared to a sequential execution
if the external concurrency model enforces execution in a single control flow.
At the same time, even if agents are internally sequential and predictable,
an external model mapping them on multiple threads may lead to unpredictable interleaving of actions, 
thus affecting predictability of the whole MAS.

We further elaborate on this in \Cref{sec:taxonomy}, 
where we present different models of external concurrency that are at the core of this contribution.


\subsection{Related Work}\label{sec:related_agent_concurrency}

The existing literature on concurrency in \ac{BDI} systems mainly focuses on
\emph{internal} concurrency.
For instance,
a recent survey~\cite{SilvaML20} provides an overview of \ac{BDI} architectures,
including considerations on how different platforms deal with the interleaving of agents' intentions.
Moreover,
the discussion about concurrency in \ac{BDI} systems 
typically concerns the \emph{interleaving} of sequentially-executed intentions,
rarely about their \emph{parallel} execution---also known as \emph{true concurrency}~\cite{Silva2020AnOS}.
Interaction among agents that need to share mutable data has also received attention.
In particular,
the shared data has been modelled with the abstraction of \emph{artifact}~\cite{ricci13-akifest},
capturing \emph{safety} and \emph{synchronisation};
adopting specialised abstractions can in turn impact internal concurrency~\cite{aloo-agere2013}.
Finally, the impact of concurrency on performance has been investigated in~\cite{ZatelliRH2015},
focussing on the effects that different concurrency configurations mapping the agent control loop can have on both  individual agents and the whole MAS.

\subsection{Underlying Concurrency Mechanisms}
\label{sec:concurrency_abstractions}

The structured programming theorem~\cite{BohmJacopini1966} states that any computable function can be expressed in terms of
selection (executing one of two subprograms depending on a condition),
iteration (repeatedly executing a subprogram until a condition is met),
and
\emph{sequence} (executing a subprogram after another).
The latter is the foundation of the so-called \emph{control flow} of a program,
and it is rooted in the assumption that instructions are \emph{totally} ordered.
In concurrent programs, instead,
the execution of instructions is rather \emph{partially} ordered~\cite{Lamport1978}:
although subprograms are executed in a given order,
instructions of different subprograms may interleave,
producing a different total ordering.
Concurrent execution can be especially beneficial
(and difficult to govern~\cite{Batty2017})
when the underlying architecture supports multiple control flows
(multiple processors, cores, or portions of the execution pipeline).

The realisation of concurrent programs
boils down to
minimising the amount of ordering constraints imposed on the execution of instructions
while guaranteeing correctness,
and
can be performed through formal or practical tools. 
Formalism dedicated to concurrent programming include process algebras~\cite{Hoare78}, \ac{CCS}~\cite{Milner80}, Petri nets~\cite{Petri1966}, and actors~\cite{Agha1986}.
From a practical perspective,
some of them are captured by programming languages,
with either a dedicated syntax or libraries,
sometimes adopting a custom naming convention,
ultimately preserving the underlying semantics.
In the following,
we introduce the most common concurrency abstractions available in most modern programming languages.

\paragraph{Threads.}

Threads are a facility provided by \acp{OS} to execute sequential programs that share memory;
they are considered the basic unit of concurrency~\cite{Dijkstra1965}.
Although the code executed by each thread is sequential,
multiple threads run concurrently
(scheduled by the \ac{OS} onto multiple logical cores and/or in a time-sharing fashion),
thus the execution of multiple threads may interleave arbitrarily.
Since they share memory,
threads may easily interact with each other by reading/writing the same memory locations,
causing race conditions and other concurrency-related issues.
Thus, multi-threaded programs commonly require synchronisation,
typically achieved by means of arguably low-level primitives
such as \emph{locks}, \emph{semaphores}, and \emph{monitors},
enforcing partial ordering among instructions of different threads.
Other concurrency abstractions are constructed by coordinating threads by means of these and similar mechanisms.

\paragraph{Processes.}

Processes are similar to threads,
but (normally, in modern \acp{OS}) they do not share memory;
rather,
inter-process communication occurs through \ac{OS} mediation via mechanisms
such as \emph{pipes}, \emph{sockets}, or the \emph{file system}.
Internally, processes can spawn multiple threads:
thus, from a concurrency perspective,
they can be intended as containers of threads sharing the process' memory space.

\paragraph{Event Loops.}

In event-driven programming~\cite{DabekZKMM2002},
event loops are abstractions to express concurrent programs while hiding the intricacies of low-level thread synchronisation.
An event loop is a single thread executing multiple tasks (subprograms) sequentially from different sources
(users, the \ac{OS}, or other parts of the program).
Tasks can be scheduled by registering the corresponding subprogram on the event loop;
internally, this operation appends the subprogram to a \ac{FIFO} queue internal to the event loop.
The event loop's thread executes the tasks in the queue in order,
waiting if the queue is empty:
any task scheduled on the event loop is \emph{eventually} executed.
The perception of parallelism of an event loop come from the fact
that new tasks can be scheduled with no need to wait for any previous one to be completed.
On the other hand,
the sequential nature of event loops becomes evident in case of long-running tasks
(e.g., I/O operations),
that may lead subsequent ones to starvation.
To mitigate this issue,
event-loops are commonly coupled with non-blocking I/O~\cite{BuettnerKL09},
where blocking read/write primitives are replaced with asynchronous events.

Notably, 
event loops are the backbone of many interesting features that are popping up in modern programming languages
-- there including JavaScript, Python, C\#, etc. --,
such as \texttt{Promise}s, \texttt{async}hronous functions, and \texttt{await} operators (cf.~\cite{LoringML17}).
In these languages, 
the event loop is hidden under the hood,
and developers are not required to interact with it directly,
but rather by means of the aforementioned features.
As far as this paper is concerned, 
we stick to the low-level abstraction of the event loop,
as our goal is to make concurrency controllable for \ac{AOP} developers%
---rather than hiding it via syntactic sugar.

\paragraph{Executors.}

Borrowing from the Java's nomenclature,\footnote{\url{https://archive.is/zF1FL}}
executors generalise event loops by supporting multiple threads.
From the user viewpoint, 
executors are essentially event loops with a configurable number of threads.
They support tasks to be enqueued in the same way as event loops do,
yet consumption of tasks from the queue is transparently performed by multiple threads
(thus, potentially, in parallel).
Executors may be further categorised depending on whether their backing thread count can change at runtime.
Fixed-sized executors are created with a specific count number of threads $N$,
which imposes an upper bound on the maximum degree of parallelism,
as at most $N$ tasks may be executed in parallel at any given moment.
Conversely, variable-sized executors may \emph{dynamically} change the number of threads in response to the runtime conditions.
A typical case where variable-sized executors are preferable is in the presence of multiple long-running blocking tasks.
For instance,
assume $N$ such tasks to be selected for parallel execution:
the fixed-sized executor would be blocked, starving the other tasks and leaving resources unused;
the variable-sized executor, instead, could spawn new threads to execute the other tasks,
and let them terminate once no blocking tasks are being run.

\subsubsection{Concurrency Abstractions in Practice.}

Although all the aforementioned concurrency abstractions are equivalent in terms of expressiveness,
there are relevant practical implications associated with any choice.

Consider, for instance, the \ac{CCS} system
$\T{a} \cdot \T{b} \cdot \T{c} \parallel \T{x} \cdot \T{y} \cdot \T{z}$,
modelling two parallel suprograms performing a sequence of atomic tasks.
Such a system, as specified, allows tasks to interleave arbitrarily,
as far as their order within the subprogram is respected
(for instance, $\T{b}$ can never happen before $\T{a}$, but $\T{a}, \T{x}, \T{y}, \T{z}, \T{b}, \T{c}$ is a perfectly valid execution).
When subprograms are executed by independent threads, this semantics is respected.
When using an event loop,
instead, some combinations become impossible,
as the execution of the next task is scheduled after the previous one's completion;
consequently,
if both $\T{a}$ and $\T{x}$ are enqueued,
only two round-robin inter-leavings are possible,
depending on which one is on the top of the queue:
$\T{a}, \T{x}, \T{b}, \T{y}, \T{c}, \T{z}$ or $\T{x}, \T{a}, \T{y}, \T{b}, \T{z}, \T{c}$.
So, we say that implementing the concurrent system on an event loop reduces the non-determinism as well as developers' degrees of control.
With an executor,
all possible interleavings are still possible,
but the degree of parallelism can be selected.

Generalising on this observation, 
we may state that the choice of concurrency abstraction has an impact on the determinism and controllability of the concurrent system.

\subsubsection{About \ac{BDI} Technologies}

Since the introduction of the \acl{BDI} framework~\cite{bratman1988bdi},
the community produced many technologies for programming \ac{BDI} agents,
most of which are based on (or inspired by) the \agentspeak{} semantics~\cite{Rao1996}.
In the remainder of this paper,
we compare several major \ac{BDI} technologies from a concurrency perspective.
We focus on those technologies that appear to have some running software implementation
that is actively maintained and used by the community.
Hence,
we build upon the recent work by Calegari \textit{et al.}~\cite{lptech4mas-jaamas35},
which surveys the state-of-the-art of logic-based agent-oriented technologies,
and we select the ones aimed to support general-purpose \ac{BDI} agents programming,
via \emph{open-source} software implementations.
Because of this criterion,
we include in our analysis the following technologies:
\astra~\cite{CollierRL15},
\goal{}~\cite{Hindriks2009},
\jadex~\cite{PokahrBL2005},
\jakta~\cite{BaiardiBCP23},
\jason~\cite{BordiniHW2007},
\phidias~\cite{DUrsoLS19},
and
\spadebdi~\cite{PalancaRCJT22}.
Of these,
\astra, \jadex, \jakta, and \jason{} are based on the \ac{JVM} platform,
\phidias{} and \spadebdi{} are based on Python,
and \goal{} is based on both the \ac{JVM} and SWI-Prolog~\cite{WielemakerHM08}%
---details about the underlying software plaftorm are relevant 
to understand the empirical analysis described in \Cref{par:inspection}.

We remark that this selection is not meant to be exhaustive.
In particular, 
many well-known \ac{AOP} technologies are not included in our analysis
(precisely because they are not \ac{BDI}),
namely: \jade{}~\cite{BellifemineCG07}, \spade{}~\cite{PalancaRCJT22}, SARL~\cite{iat2014}, or Kiko~\cite{ChristieSC2023}.
Similarly,
we exclude \ac{BDI} technologies that are closed source
-- e.g., \jack~\cite{Winikoff2005} --,
or not actively maintained 
(at the time when the survey by Calegari \textit{et al.} \cite{lptech4mas-jaamas35} was conducted)%
---e.g., AFAPL\footnote{\url{https://archive.ph/Hozbl}}, 
2APL~\cite{Dastani08},
or 3APL~\cite{HindriksBHM99}.

\section{A Taxonomy of Concurrency Patterns for \ac{MAS} Execution}
\label{sec:taxonomy}

In this section we identify the most relevant \emph{external} concurrency models for \acp{MAS}---namely, how the atomic parts of the agent' control loop get mapped onto the underlying abstractions
described in \Cref{sec:concurrency_abstractions}.
Different \emph{internal} concurrency models dictate different levels of granularity of the atomic components of the control loop,
thus influencing the \emph{external} concurrency model.
Consequently, 
we focus on the mapping of the largest possible autonomous unit in \ac{AOP}, the entire agent,
discussing the potential external concurrency models for a \ac{MAS}.

\paragraph{\ac{1A1T}.}

Each agent is mapped onto a single thread, which executes its whole control loop.
Hence, the \ac{MAS} consist of several threads managed by the \ac{OS} scheduler,
and the interleaving among different agents' operations is unpredictable.
The controllability of the \ac{MAS} execution is abysmal, as control is delegated to the \ac{OS};
for the same reason, determinism is minimal.
Additionally, with \ac{1A1T} the active thread count of the \ac{MAS} is unbound,
and when such count largely exceeds the logical processors performance degrades~\cite{DBLP:journals/sigops/LingML00}.

\paragraph{\ac{AA1T}.}

The whole \ac{MAS} is executed on a single thread which internally schedules all agents' execution in a custom way,
following some (usually cooperative) scheduling policy---e.g., round-robin.
Most commonly, 
the internal scheduling policy alternates agents at the control loop step or stage level.
By omitting the former case for the sake of conciseness, the latter can be formalised as follows (provided that the scheduling policy is round-robin):
\begin{equation}\label{eq:aaoel}
 \begin{array}{rcl}
   \V{mas} & \defeq & \V{sense} \cdot \V{deliberate} \cdot \V{act} \cdot \V{mas} \\
   \V{sense} & \defeq & \T{sense}_1 \cdot \ldots \cdot \T{sense}_N \\
   \V{deliberate} & \defeq & \T{deliberate}_1 \cdot \ldots \cdot \T{deliberate}_N \\
   \V{act} & \defeq & \T{act}_1 \cdot \ldots \cdot \T{act}_N \\
 \end{array}
\end{equation}
Using a single thread with custom internal scheduling policy renounces parallelism (hence, performance) in
favour of controllability:
thus, it is a good choice 
when determinism, reproducibility, and predictability are primary concerns---as in many simulated or time-critical scenarios. 
Notably,
because of the cooperative nature of the scheduling,
sensing, deliberation, and actuation operations should terminate as quickly as possible,
to avoid blocking the whole \ac{MAS}.

\paragraph{\ac{AA1EL}.}

The whole \ac{MAS} is executed on a single event loop,
which internally schedules all agents' execution with a FIFO queue of tasks,
ensuring fairness by design if all new tasks in the event loop are inserted by other tasks of the event loop.
At the conceptual level,
this is equivalent to a fair \ac{AA1T};
as such,
controllability and performance are akin to \ac{AA1T}.
In practice, however,
\ac{AA1EL} requires explicitly modelling the agent's control loops activities as tasks on the event loop,
thus, despite the conceptual equivalence,
technical implementations of \ac{AA1T} and \ac{AA1EL} may be fairly different.

\paragraph{\ac{AA1E}.}
Similarly to \ac{AA1EL},
each atomic operation is mapped onto a task to be enqueued on a shared executor.
This model enables the parallel execution of multiple agents, and,
if the internal concurrency model supports it,
the parallel execution of the same agent's activities.
In case each agent enqueues at most one task at a time
(a solution often used to enforce consistency),
\ac{AA1E} is conceptually equivalent to \ac{1A1T}.
However, \ac{AA1E} is preferable from a technological perspective,
as the agent (and agent's actions) count is decoupled from the thread count,
resulting in finer control on the degree of parallelism
(by governing the amount of threads in the executor),
as well as in a better exploitation of the underlying resources.
Furthermore, 
the granularity of the interleaving may be tuned by choosing how to model events and tasks.

For instance, 
one may model the entire control loop step as a single task,
hence making the interleaving more coarse-grained%
---formally:
\begin{equation*}
 \begin{array}{rcl}
   \V{mas} & \defeq & \V{agent}_1 \parallel \ldots \parallel \V{agent}_n \\
   \V{agent} & \defeq & \V{step} \cdot \V{agent} \\
   \V{step} & \defeq & \T{step} \\
 \end{array}
\end{equation*}
where $\T{step}$ represents one single execution of all stages, from sensing to acting.

Two further specialisations of this model are possible, 
depending on whether the executor is fixed- or variable-size.
In the former case, 
there is an upper bound on the amount of threads the \ac{MAS} can leverage.
Although helpful to limit the resource exploitation in constrained environments,
it may introduce subtle interdependencies among agents.
For instance,
when there are $M$ agents and $N<M$ threads,
if $N$ agents are performing blocking operations,
then the other $M-N$ agents must wait.
Of course,
the fixed-sized executor with $N = 1$ is equivalent to \ac{AA1EL}.
If the executor is variable-sized,
then the number of threads is adjusted dynamically,
upon need---i.e., by trying to match the count of active threads and logical processors.
\ac{AA1E} is generally
preferable over \ac{1A1T},
as the total thread count is controllable.

\subsection{Concurrency at different levels of granularity}
\label{sec:holoinc-perspective}

Concurrency abstractions may be combined to form more complex ones.
For instance, both processes and executors are composed by threads.
Threads in a process may be part of the same executor, or multiple ones.
In a distributed setting, a system may be composed by many processes spread across a several machines,
losing shared memory and thus requiring serialisation to communicate.
In orchestration frameworks,
the same service may consist of multiple containers,
distributed on different machines,
each one running multiple processes.

In principle,
when implementing a \ac{MAS}, agents may be mapped onto any of these concurrency abstractions
with different trade-offs between flexibility and controllability.
For instance, when \ac{1A1P} is adopted,
the agent's internal control loop may be implemented with multiple threads,
but communication among agents will require (de)serialisation,
as agents will not share memory with each other.
For \ac{BDI} agents,
threads may be used to model intentions,
paying a price in terms of implementation complexity
(as agent-specific synchronisation mechanisms would be required)
to obtain an extremely fine-grained degree of control over the execution.

Combinations (and complexity) can scale arbitrarily,
as in principle any \ac{AOP} abstraction can be mapped on any lower-level concurrency abstraction%
---thus allowing uncommon combinations such as One-Agent-One-Container or One-Intent-One-Process.
For the sake of simplicity,
in this paper we focus on the cases listed in \Cref{sec:taxonomy},
which we show suffice to capture the behaviour of all the selected \ac{BDI} technologies.
However,
we discuss the implications of more nuanced concurrency models briefly in \Cref{sec:discussion}.

\section{Analysis of BDI Technologies and Concurrency Models}\label{sec:comparison}

In this section,
we inspect the external concurrency models supported by a selection of
actively-maintained open source \ac{BDI} programming frameworks.
In particular, we focus on
\astra~\cite{CollierRL15},
\goal{}~\cite{Hindriks2009},
\jadex~\cite{PokahrBL2005},
\jakta~\cite{BaiardiBCP23},
\jason~\cite{BordiniHW2007},
\phidias~\cite{DUrsoLS19},
and
\spadebdi~\cite{PalancaRCJT22}.
We do not claim this selection to be exhaustive, so we leave a more complete analysis for future works.

\subsection{Methodology}

We performed our analysis in three steps:
\begin{enumerate}
  \item \emph{empirical evaluation} through a synthetic benchmark
    designed to
    reveal how many threads are involved in the execution of a \ac{MAS} and how they interleave;
  \item \emph{documentation and source code inspection}
    to understand implementation details and customisability.
  \item \emph{direct contact} with the current maintainers,
    asking for confirmation of our findings and for further details,
    including a subjective evaluation of the feasibility of supporting additional external concurrency models.
\end{enumerate}

\subsubsection{Empirical Evaluation.}\label{par:benchmark}

We created a benchmark~\cite{Baiardi_BDI_Languages_Concurrency_2024} to reveal how threads are leveraged in a \ac{BDI} \ac{MAS}.
The benchmark consists of a simple \ac{MAS}, composed by two agents enacting one round a ping--pong protocol:
the pinger agent initiates the protocol by sending a message to the ponger agent,
which replies by sending the a message back to the pinger.
To reveal how threads are used,
we make agents execute a custom action
-- \texttt{revealCurrentThread} --
before and after each message sending and reception.
As a reference, we show our \jason{} implementation for
pinger (\Cref{lst:pinger_example}) and ponger (\Cref{lst:ponger_example}).
To maximise the likelihood of intercepting all threads,
when supported
we force agents to pursue different intentions simultaneously
(in the reference specification, this is done through the \jason{} operator \texttt{!!}).
\begin{lstlisting}[
  float,
  linewidth=\linewidth,
  label={lst:pinger_example},
  caption={ASL description for \emph{pinger} agent.}
]
!ping.

+!ping <- 
    .revealCurrentThread("intention 1");
    .send(pong, tell, ball);
    !!showThread(2);
    .revealCurrentThread("intention 1").

+ball <-
    !!showThread(4);
    .revealCurrentThread("intention 3").

+!showThread(X) <- .revealCurrentThread("intention " + X).
\end{lstlisting}
\begin{lstlisting}[
  float,
  linewidth=\linewidth,
  label={lst:ponger_example},
  caption={ASL description for \emph{ponger} agent.}
]
+ball[source(X)] <-
    .revealCurrentThread("intention 5");
    .send(X, tell, ball);
    !!showThread(6);
    .revealCurrentThread("intention 5").

+!showThread(X) <- .revealCurrentThread("intention " + X).
\end{lstlisting}
We then analyse the trace obtained
by multiple executions of the benchmark,
and we use it to understand
how many threads the agent is using to execute its intentions
and in which order intentions (of different agents) are executed and interleaved.
By repeating this analysis we empirically infer which concurrency model is used to execute agents.

One possible trace log is reported in \Cref{lst:test_output_jason}, 
showing the benchmark output when executed in \jason{}%
---as implemented in \Cref{lst:pinger_example} and \ref{lst:ponger_example}.
\begin{lstlisting}[
 float,
 linewidth=\linewidth,
 label={lst:test_output_jason},
 caption={An example of execution on which each agent executes its steps on its thread.}
]
[pinger] Intention 1 is executed on thread: pinger
[pinger] Intention 1 is executed on thread: pinger
[ponger] Intention 5 is executed on thread: ponger
[pinger] Intention 2 is executed on thread: pinger
[ponger] Intention 5 is executed on thread: ponger
[pinger] Intention 3 is executed on thread: pinger
[ponger] Intention 6 is executed on thread: ponger
[pinger] Intention 4 is executed on thread: pinger
\end{lstlisting}
There, logs suggest that \jason{} may implement \ac{1A1T} concurrency model, as each agent is always logged by the same thread, the same thread is never used by two different agents, and the thread identifier is associated to the name assigned to the agent in the \ac{MAS} specification.
Accordingly, for each \ac{BDI} technology, we re-implement the benchmark in the most idiomatic way possible, and we run it.

We were not able to reproduce our benchmark on \jadex{}, \spadebdi{}, and \goal{}.
We then did not infer the concurrency model used in these technologies
and moved to the following step of our evaluation.

\subsubsection{Documentation and Source Code Inspection.}\label{par:inspection}

In general, the empirical evaluation can let \emph{some} external concurrency model emerge,
but it cannot be exhaustive:
as discussed in \Cref{sec:concurrency_abstractions},
some abstraction may not show all their possible behaviours even after repeated executions,
and some may produce the same outputs.
Also, 
the empirical evaluation of some platforms is more difficult to implement and less revealing.
For instance,
the \ac{JVM} thread inspection primitives  
(with which \jason{} can interact)
are more expressive than SWI-Prolog ones
(with which \goal{} interacts).
We thus inspect the source code and the official documentation of the surveyed frameworks
to learn as much as possible.

In detail, 
while inspecting the source code of \ac{JVM}-based \ac{BDI} technologies, 
we look for usages of (Java) standard-library classes such as \texttt{Thread}, \texttt{Executor}, \texttt{ExecutorService}, \texttt{ForkJoinPool}, 
as well as usages of parallel streams.
Similarly,
for Python-based technologies, 
we look for usages of (Python) standard-library types such as \texttt{Thread}, \texttt{Abstract\-EventLoop} (or subclasses), \texttt{Task}, 
as well as usages of coroutines.

After inferring the concurrency model of each \ac{BDI} technology, 
we assess if and to what extent the concurrency model can be \emph{customised} by the users of that technology,
by looking for ad-hoc syntax or \ac{API} letting \ac{MAS} specifications customise the concurrecy model.

\subsubsection{Direct Contact.}
Once the results from the previous steps were gathered,
we contacted maintainers of each framework 
so as to confirm our assessment and gain additional insights.
This operation was useful to get past what is available out-of-the-box,
and what could be achieved with reasonably limited extensions.
We described the developers the taxonomy of \Cref{sec:taxonomy}
and reported our results.
We asked them to evaluate on our findings,
adding comments about whether the non-supported external concurrency models were
available out of the box (thus, missed by the analysis),
could be supported with reasonable effort,
or required extensive rewriting of the codebase: 
a template of the email sent to all maintainers can be found in \Cref{appendix:questions}.
We received answers from all developers except for \jason{} and \spadebdi{};
all answers confirmed our initial results.

\subsection{Results}
\label{subsec:results}

\Cref{tab:experiment_results} summarises the results of our analysis.
When evaluating \ac{1A1P}, we also require agents to be capable of inter-process communication,
e.g., by means of protocols such as TCP/IP\@.
In the rest of this section,
we detail how the analysis is performed for each technology,
summarising the most prominent findings.

\begin{table}[!tb]
  \centering
  \caption{
    Summary of \ac{BDI} technologies and their concurrency models.
    The symbols $\checkmark$, $\thicksim$, and $\times$ indicate, respectively,
    that the concurrency model is supported, that it could be supported with a custom implementation, and that it is not supported.
  }
  \label{tab:experiment_results}
    \begin{tabular}{r||c|c|c|c|c|c}
      \toprule
      \textbf{Model} $\rightarrow$ & \textbf{\ac{1A1T}}  & \textbf{\ac{AA1T}} & \textbf{\ac{AA1EL}} & \textbf{\ac{AA1E}}  & \textbf{\ac{AA1E}} & \textbf{\ac{1A1P}} \\
      \textbf{Tech.} $\downarrow$ & & & & \textbf{fixed}  & \textbf{variable} \\
      \midrule
      \textbf{\astra} & $\thicksim$ & $\thicksim$ & $\checkmark$ & $\checkmark$ & $\checkmark$ & $\thicksim$ \\
      \textbf{\goal} & $\checkmark$ & $\times$ & $\times$ & $\times$ & $\times$ & $\times$ \\
      \textbf{\jadex} & $\thicksim$ & $\checkmark$ & $\thicksim$ & $\thicksim$ & $\checkmark$ & $\checkmark$ \\
      \textbf{\jakta} & $\checkmark$ & $\checkmark$ & $\checkmark$ & $\checkmark$ & $\checkmark$ & $\thicksim$ \\
      \textbf{\jason} & $\checkmark$ & $\thicksim$ & $\checkmark$ & $\checkmark$ & $\thicksim$ & $\checkmark$ \\
      \textbf{\phidias} & $\checkmark$ & $\times$ & $\times$ & $\times$ & $\times$ & $\checkmark$ \\
      \textbf{\spadebdi} & $\times$ & $\times$ & $\checkmark$ & $\times$ & $\times$ & $\checkmark$ \\
      \bottomrule
    \end{tabular}
\end{table}

\subsubsection{\astra{}.}

\astra{}~\cite{CollierRL15} is a \ac{BDI} agent technology written in Java
designed with a C-family syntax.
\astra{} provides fine-grained control over the execution of the \ac{MAS} entities,
our benchmark indeed revealed that any iteration step of the control loop of the same agent may run on a different thread,
suggesting a \ac{AA1E} model.
Source code inspection confirmed the analysis and revealed that the executor is \emph{variable-sized}.
Since Java executors can be used as event loops, \ac{AA1EL} is supported, too.
\begin{lstlisting}[
  float,
  linewidth=\linewidth,
  language=Java,
  label={lst:astra_executor},
  caption={Snippet of \astra{}'s code base (\url{https://gitlab.com/astra-language/astra-core}) showing how the \acl{AA1E} model is implemented.}
]
// astra-interpreter/src/main/java/astra/execution/BasicSchedulerStrategy.java
import java.util.concurrent.ExecutorService;
import java.util.concurrent.Executors;
// ...
public class BasicSchedulerStrategy implements SchedulerStrategy {
  private ExecutorService executor =
      Executors.newFixedThreadPool(2);
  // ...
  public void setThreadPoolSize(int size) { /*...*/ }
  public void schedule(final Agent agent) {
    executor.submit(() -> {
      // ...
      agent.execute(); // run one step of the agent's control loop
      // ...
      schedule(agent); // schedules the next step
    });
  }
}
\end{lstlisting}
Among the available implementations which are present in the \astra{} codebase,
the \texttt{BasicSchedulerStrategy} (see \Cref{lst:astra_executor}) 
is the one that supports the fixed-size \ac{AA1E} concurrency model.

\astra{} also supports custom implementations of the concurrency model,
which users may provide by implementing \texttt{SchedulerStrategy} interface.
Thus, \ac{1A1T} and \ac{AA1T}
could be implemented quite easily.
The maintainers of \astra{} confirmed our previous assertions,
thus describing to us that there is also the possibility to implement the \ac{1A1P} concurrency model
by the means of a custom distributed messaging infrastructure implementation for \astra{} \texttt{Messaging} class,
that currently is not provided.

\subsubsection{\goal{}.}

\goal{}~\cite{Hindriks2009} is a Java \ac{BDI} library distributed as an Eclipse IDE plugin
that integrates with SWI Prolog through \ac{JNI}~\cite{DBLP:journals/crossroads/Husaini97},
that does not expose Java primitives.
Due to its peculiar integration with SWI Prolog~\cite{WielemakerHM08},
\goal{} is bound to the \ac{1A1T} model,
and it does not support customisations
without major changes to the code base.
However,
the library comes with an option for emulating \ac{AA1T};
although internally agents are still executed on different threads,
these are executed sequentially.
The maintainers of \goal{} also confirmed that, 
due to this strict integration with SWI Prolog,
other concurrency models cannot be custom-built,
even with major code changes.

\subsubsection{\jadex{}.}

\jadex{}~\cite{PokahrBL2005} is a \ac{BDI} Java library.
We analysed the latest version of the library, namely \jadex{} V,
which improved modularisation
and simplified agents' concurrency management through (\jadexV{}'s terminology)
\texttt{ExecutionFeature}s.
\jadexV{} natively supports variable-sized \ac{AA1E} as default behaviour, \ac{1A1P} and \ac{AA1T}.
Further customisation options could be implemented with custom \texttt{ExecutionFeature}s modules.

\subsubsection{\jakta{}.}

\jakta{}~\cite{BaiardiBCP23} is a Kotlin-based \ac{DSL}\footnote{
    \url{https://archive.is/El3fE}
} for \ac{BDI} \ac{MAS} running on the \ac{JVM}.
It exposes the concurrency model as a first-class abstraction 
in the \ac{DSL} (see \Cref{lst:jakta_configuration})
supporting custom models
through the implementation of the \texttt{ExecutionStrategy} interface.
\begin{lstlisting}[
    float,
    linewidth=\linewidth,
    language=Kotlin,
    label={lst:jakta_configuration},
    caption={Example of \ac{MAS} configuration with execution strategy customisation in \jakta{}.}
]
mas {
    environment { /* external actions' definitions here */}
    agent("pinger") { /* pinger specification here */ }
    agent("ponger") { /* ponger specification here */ }
    executionStrategy { oneThreadPerAgent() } // first-class support
}.start()
\end{lstlisting}
By default, \jakta{} uses \ac{AA1T}
to support reproducibility while debugging or simulating;
but \ac{1A1T}, \ac{AA1E}, and \ac{AA1EL}
are also supported natively.
\ac{1A1P} is not supported out of the box,
but could be implemented through an extension.

\subsubsection{\jason{}.}

\jason{}~\cite{BordiniHW2007} is a well-known
\agentspeak{}-compliant
\ac{BDI} agent technology implemented in Java.
\jason{} defaults to a \ac{1A1T} model,
yet the concurrency models are customisable
by specifying a different \emph{infrastructure}.
Similarly to \jakta{},
these can be configured at \ac{MAS} specification time and customised;
implementations available out of the box provide support for
\ac{AA1T} (\texttt{Local/threaded} infrastructure),
fixed-sized \ac{AA1E} and \ac{AA1EL} (\texttt{Local/pool} infrastructure);
and \ac{1A1P} (\texttt{Jade} infrastructure).

\subsubsection{\phidias{}.}

\phidias{}~\cite{DUrsoLS19} is a Python internal \ac{DSL}
defaulting to the \ac{1A1T} concurrency model.
We reach this conclusion by code inspection: 
Python's threads are explicitly created behind the scenes of each agent%
---as shown in \Cref{lst:phidias_thread_creation}, 
which is taken from \phidias{}' codebase.
Even using threads,
the execution in most Python interpreters is not parallel
because of the \ac{GIL}.\footnote{\url{https://archive.is/5KBqn}}

Inter-agent communication is implemented through HTTP,
suggesting that \ac{1A1P} is supported too.
However, there is no way to customise the concurrency model.
In other words, the \ac{1A1T} and \ac{1A1P} models are hard-coded.
The maintainers of \phidias{} validated our analysis,
confirming that only an entire model redesign would enable other concurrency models.

\begin{lstlisting}[
 float,
 linewidth=\linewidth,
 language=Python,
 label={lst:phidias_thread_creation},
 caption={Snippet of \phidias{} code base (\url{https://github.com/corradosantoro/phidias}) showing how agents are executed on threads.}
]
# phidias/lib/phidias/Runtime.py
import threading
class Runtime:
    # ...
    @classmethod
    def run_agent(cls, a):
        e = cls.engines[a]
        t = threading.Thread(target=e.run)
        t.start()
\end{lstlisting}

\subsubsection{\spadebdi{}.}

\spadebdi{}~\cite{PalancaRCJT22} is a Python library adding \ac{BDI} agents on top of \spade{}~\cite{GregoriCB06}.
In turn, 
\spade{} supports inter-agent communication by means of the XMPP protocol, 
and a notion of agent similar to \jade{} (cf.~\cite{BellifemineCG07}).

\begin{lstlisting}[
 float,
 linewidth=\linewidth,
 language=Python,
 label={lst:spadebdi_ag_execution},
 caption={
  Agent execution in \spadebdi{} inherits \spade's event loops, 
  which in turns are a feature from Python standard library,
  as demonstrated by these code snippets from the \spadebdi{} 
  (\url{https://github.com/javipalanca/spade_bdi}) 
  and \spade{} (\url{https://github.com/javipalanca/spade}) codebases.
 }
]
# spade_bdi/bdi.py
from spade.agent import Agent
class BDIAgent(Agent):
   # ...

# spade/agent.py
from spade.container import Container
class Agent(object):
    def __init__(self, ...):
        # ...
        self.container = Container()
        # ...
        self.loop = self.container.loop
        # ...

# spade/container.py
class Container(object):
    def __init__(self):
        # ...
        self.loop = get_or_create_eventloop() # uses Python's asyncio API
        # ...
    
    def run(self, coro: Awaitable) -> None:
        self.loop.run_until_complete(coro)
\end{lstlisting}
Agents in both \spadebdi{} and \spade{} are implemented via Python's native event loops and coroutines,
providing native support for \ac{AA1EL}.
In particular, 
the \texttt{BDIAgent} class extends \spade{}'s \texttt{Agent} one, 
which in turns models the agent's control loop as a coroutine.
As shown in \Cref{lst:spadebdi_ag_execution} 
-- taken from \spade{}'s codebase --, 
coroutines are executed on event loops via standard Python \ac{API}.
We infer that \spadebdi{} supports the \ac{AA1EL} and possibly the \ac{1A1P} concurrency models 
-- as agents may interact across different processes via XMPP --, 
whereas customisability is not an option---as far as documented \ac{API} are concerned.

\section{Conclusion, Recommendations, and Future Works}\label{sec:discussion}

The external concurrency model is a key aspect to be taken into account when designing or using a (\ac{BDI}) \ac{MAS} technology.
Generally speaking,
the more options the better,
as applications can be finely tailored to the specific requirements of the problem at hand.

\paragraph{Controllability.}

Reproducibility and controllability are key aspects of \ac{MAS} engineering,
especially during debug and simulation.
When full control is required,
\ac{AA1T} and \ac{AA1EL} are the best choices,
as they enforce a single control flow.

\paragraph{Performance.}

Better Performance is generally achieved by exploiting parallelism.
Although the \ac{1A1T} model seems attractive for its simplicity,
\ac{AA1E} is preferable,
especially whenever the agent count largely exceeds the logical processors.

\paragraph{Design of \ac{BDI} Technologies.}

We argue that designers of \ac{BDI} technologies should provide means to customise the concurrency model
with a dedicated and well-documented \ac{API}.
Doing so requires careful consideration of the concurrency model as early as possible.
Building a \ac{BDI} platform around assumptions on the desired concurrency model may simplify the implementation,
but it is likely to backfire later on,
limiting extensibility and applicability%
---for instance, by preventing the system to scale up and down depending on the available resources.
If assumptions are to be made due to technical constraints,
some choices are more flexible than others;
for instance, \ac{AA1E} can emulate \ac{1A1T} and \ac{AA1EL},
while the opposite does not apply.
The key to design \ac{BDI} platforms capable to adapt to multiple concurrency models is the
\emph{complete separation between agents' control loop and their target concurrency abstraction}.
Following this principle,
it should be possible to write the \ac{MAS} specification once,
then run it unchanged on different concurrency models.
\paragraph{Impact on Internal Concurrency.}

We discuss how internal concurrency models impact external concurrency models
by bounding the maximum \emph{granularity} at which the agent's control loop can be parallelised.
However, the influence is bidirectional:
enforcing external abstractions binding specific \ac{BDI} abstractions to one or more control flows
(such as \ac{1A1T} or \ac{AA1T}) may hinder further attempts to control the degree of parallelism
by exploiting finer-grained internal concurrency models
(e.g., parallelise at the level of intentions).

\paragraph{Final Remarks.}

The external concurrency of \ac{BDI} agents is a paramount aspect of practical \ac{MAS} engineering.
In this paper,
we define clear terminology and taxonomy to support decision-making about concurrency in \ac{MAS},
addressing both the construction of \ac{MAS}
and the (re)design of \ac{BDI} technologies.
We analyse the state of the art of several relevant \ac{BDI} technologies,
showing that there is heterogeneity in terms of supported concurrency models and their customisability.
We advocate for further research efforts
to provide \ac{BDI} technology designers with clear guidelines and best practices
regarding practical external concurrency models,
favouring harmonisation and standardisation.

\subsection{Current Limitations and Future Works}

One limitation of our work is that 
we focus on relatively small set of well-established \ac{BDI} technologies:
this is a deliberate choice of ours.
Our intent,
in this work,
is to give a concise, self-contained, and insightful overview of the technical issues
arising from the external concurrency of \ac{MAS} technologies.
However,
we acknowledge that this work is just the first step in this direction.
Along this line,
we plan to apply our inspection on a wider range of \ac{MAS} technologies,
possibly adopting exhaustiveness as a criterion.

In fact, 
it is worth mentioning that our inspection methodology could be applicable,
in principle,
to \emph{any} \ac{AOP} technology,
regardless of whether it is \ac{BDI} or not:
\emph{external} concurrency is a key aspect of \ac{AOP} technologies in general,
whereas \emph{internal} concurrency 
-- as it is defined in this paper --
is specific for the \ac{BDI} paradigm%
---and, specifically, to the notion of \emph{intention}.

Accordingly,
we plan to extend our analysis to other \ac{AOP} technologies,
possibly beyond the realm of \ac{BDI} architectures.
Along this line,
we also plan to widen the definition of \emph{internal} concurrency,
to account for other behavioural abstractions, 
possibly adopted by other \ac{AOP} architectures.
This would be for instance the case of \emph{behaviours} in \jade~\cite{BellifemineCG07} or \spade~\cite{PalancaRCJT22}.

Another limitation of our work is that 
we do not really explore the relationship among concurrency models and agents' \emph{interaction}.
While it may be reasonable,
for a first exploration of the topic,
to consider agent's interaction as a by-product of agent's perceptions and actions
-- and therefore negligible in terms of concurrency --,
we acknowledge that this is a simplification.
In particular, 
this may be too simplistic when \ac{AOP} technologies are applied to distributed systems
(where message passing cannot be reduced to an atomic operation),
or when the focus of the \ac{MAS} technology of choice is interaction itself
(e.g. communication protocols).

Accordingly,
we plan to extend our analysis to the interaction dimension of \ac{MAS}.
There, 
we intend to study the interplay among concurrency models and interaction patterns.
Along this line,
including non-\ac{BDI} technologies tailored on interaction
(cf. Kiko~\cite{ChristieSC2023})
would be paramount.

\section*{Acknowledgements}

This work has been partially supported by:
\begin{inlinelist}
    \item ``WOOD4.0 - Woodworking Machines for Industry 4.0'',
        Emilia-Romagna CUP E69J22007520009;
    \item ``FAIR--Future Artificial Intelligence Research'',
        Spoke 8 ``Pervasive AI''
        (PNRR, M4C2, Investimento 1.3, Partenariato Esteso PE00000013),
        funded by the EC under the NextGenerationEU programme;
    \item “ENGINES — ENGineering INtElligent Systems around intelligent agent technologies” project funded by the Italian MUR program ``PRIN 2022'' (G.A.\ 20229ZXBZM),
    and
    \item 2023 PhD scholarship
        (PNRR M4C2, Investimento 3.3 DM 352/2022),
        co-funded by the European Commission
        and AUSL della Romagna.
\end{inlinelist}
Also, the authors would like to thank all the researchers and developers who answered our request for comments for their invaluable help.

\appendix

\section{Appendix: Framework Maintainers Interview}
\label{appendix:questions}

\begin{lstlisting}[breaklines, breakatwhitespace=true, frame=none, keepspaces=false, columns=flexible, tabsize=2]
Dear <Maintainer>,
we are reaching out to you to ask information about <X>.

Our research group is conducting research on how MAS platforms deal with the underlying concurrency mechanisms.
We are surveying several technologies to understand how they map the agents' lifecycle on the underlying mechanisms:

  1. One-Agent-One-Thread: Each agent is mapped into a single thread.
  2. All-Agents-One-Thread: The whole MAS is executed on a single thread, following a scheduling policy (i.e. Round-Robin).
  3. All-Agents-One-Event-Loop: The MAS is executed over an event-loop.
  4. All-Agents-One-Executor: Similar to case 3, but it uses threads to allocate agents, resulting in an effectively parallel execution. We distinguish two sub-cases:
      a. fixed-sized executors (static thread count)
      b. variable-sized (dynamically changing thread count).
  5. One-Agent-One-Process: which, internally, could exploit all the above taxonomies to execute its control loop.

We inspected your code source and identified that <X> currently supports <list of supported>,
however, we were not able to infer if it can supports <list of not supported>
Would you agree with the previous assertion?

Would it be possible to write custom extensions to implement <list of not supported> with no changes to the current code base of <X>?
If not, what about implementing the missing mechanisms directly?
Would it be feasible, in your opinion?
And if so, would you consider it easy, moderate, hard, or very hard?
\end{lstlisting}

\bibliographystyle{splncs04}
\bibliography{paper-2024-emas-concurrency}

\end{document}